\documentclass[useAMS,usenatbib]{mn2e}
\usepackage{amsmath}
\usepackage{graphicx}
\usepackage{txfonts}
\usepackage{natbib}
\usepackage{bm}

\def\beq#1{\begin{equation}\label{#1}}
\def\eeq{\end{equation}}
\def\beqa#1{\begin{eqnarray}\label{#1}}
\def\eeqa{\end{eqnarray}}

\def\Eq#1{Eq.~(\ref{#1})} 
\def\eqn#1{~(\ref{#1})}
\def\myfrac#1#2{\left(\frac{#1}{#2}\right)}

\def\comment#1{\relax}

\title[On properties of MRI]{On properties of Velikhov-Chandrasekhar MRI in ideal
and non-ideal plasma}
\author[N. Shakura \& K. Postnov] {N. Shakura
\thanks{E-mail: nikolai.shakura@gmail.com, kpostnov@gmail.com},
K. Postnov\\
Sternberg Astronomical Institute, Moscow M.V. Lomonosov State University, Universitetskij pr., 13, 119992, Moscow, Russia}	

\begin{document}

\date{Received ... Accepted ...}
\pagerange{\pageref{firstpage}--\pageref{lastpage}} \pubyear{2012}

\maketitle

\label{firstpage}

\begin{abstract}
Conditions of Velikhov-Chandrasekhar magneto-rotational instability in 
ideal and non-ideal plasmas are examined. 
Linear WKB analysis of hydromagnetic axially symmetric flows 
shows that in the Rayleigh-unstable hydrodynamic case 
where the angular momentum decreases with radius, the MRI branch becomes stable, 
and the magnetic field suppresses the Rayleigh instability at small wavelengths. 
We investigate the limiting transition from hydromagnetic flows 
to hydrodynamic flows. The Rayleigh mode smoothly transits to
the hydrodynamic case, while the Velikhov-Chandrasekhar MRI mode completely 
disappears without the magnetic field. 
The effects of viscosity and magnetic diffusivity in plasma on 
the MRI conditions in thin accretion discs are studied.
We find the limits on the mean free-path of ions allowing MRI to 
operate in such discs.

\end{abstract}


\section{Introduction}
\label{intro}
 
In the end of the 1950s -- beginning of the 1960s, E. Velikhov and S. Chandrasekhar
studied the stability of sheared hydromagnetic flows \citep{Velikhov59, 1960PNAS...46..253C}.
In these papers, the magneto-rotational instability (MRI) in axisymmetric 
flows with magnetic field was discovered. 
MRI arises when a relatively small seed poloidal magnetic field
is present in the fluid.
This instability was applied to astrophysical 
accretion discs in the influential paper by \citealt{1991ApJ...376..214B}, and since then
has been considered as the major reason for the turbulence arising in accretion discs 
(see \cite{1998RvMP...70....1B} for a review). Non-linear numerical simulations 
(e.g. \citealt{1995ApJ...440..742H, 2012ApJ...749..189S, 2013ApJ...772..102H}) confirmed 
that MRI can sustain turbulence and dynamo in accretion discs. However, semi-analytical and  
numerical simulations (see, for example, \citealt{2008ApJ...689.1234M, 2011IAUS..274..422S, 2013ApJ...772..102H, 2014ApJ...784..121S, 2014arXiv1409.2442N})
suggest that the total 
(Reynolds $+$ Maxwell) stresses due to MRI are insufficient to cause 
the effective angular momentum transfer in accretion discs in terms of the phenomenological 
alpha-parameter $\alpha_{SS}$ \citep{1973A&A....24..337S}, giving rather low values 
$\alpha_{SS}\sim 0.01-0.03$. Note that from the observational point of view, the alpha-parameter 
can be reliably evaluated, e.g., from the analysis of non-stationary accretion discs in 
X-ray novae \citep{2008A&A...491..267S}, dwarf-nova and AM CVn stars 
\citep{2012A&A...545A.115K}, and turns out to be an order of magnitude higher than 
typically found in the numerical MRI simulations. 

In this paper we use the local 
linear analysis of MRI in the WKB-approximation by Balbus and Hawley (1991)
to examine properties of MRI for different laws of differential rotation in weakly magnetized 
flows, $\Omega^2(r)\propto r^{-n}$, i.e. when 
the solution to linearized MHD equations in the Boussinesq approximation is searched for
in the form $\sim e^{i(\omega t-k_r r-k_z z)}$, where $k_r, k_z$ are wave vectors in
the radial and normal direction to the disc plane, respectively, in the cylindrical coordinates. 

In this approximation, the dispersion relation represents a biquadratic algebraic equation. 
The linear local analysis of unstable modes in this case was performed earlier (see, e.g., \cite{2012MNRAS.423L..50B}). Here we emphasize the different behaviour of stable and unstable modes of this equation for different rotation laws of the fluid. We show that  
in the Rayleigh-unstable hydrodynamic case 
where the angular momentum decreases with radius, the Velikhov-Chandrasekhar 
MRI does not arise, and the magnetic field suppresses the Rayleigh instability at small wavelengths.  

Then we turn to the analysis of non-ideal plasma characterized by 
non-zero kinematic viscosity $\nu$ and magnetic diffusivity $\eta$. 
This problem has been addressed previously by different authors 
(see, e.g. \cite{1998RvMP...70....1B, 
1999ApJ...515..776S, 2001MNRAS.325L...1J, 2004ApJ...616..857B, 2005ApJ...633..328I, 2008ApJ...684..498P}, among others), 
aimed at studying various aspects of the MRI physics and applications. To keep the paper self-contained, 
we re-derive the basic dispersion relation in the general case 
and investigate its behaviour for
different values of the magnetic Prandtl number $\mathrm{P_m}=\nu/\eta$ and the kinematic vicosity 
$\nu$. Specifically, we consider the limitations implied by the 
viscosity in accretion discs with finite thickness, and find phenomenologically interesting 
constraints on the disc parameters where MRI can operate.

The structure of the paper is as follows. In Section 2 we repeat the linear WKB analysis for
small perturbations in an ideal fluid and consider five different cases for MRI and Rayleigh 
modes. We also investigate the behaviour of MRI at vanishing magnetic field. 
In Section 3 we generalize the linear analysis for non-ideal plasma with 
non-zero viscosity and magnetic diffusion. First we analytically 
investigate the growth of linear perturbations in a plasma with the Prandtl number $\mathrm P_m=1$, and then consider the case of a plasma with arbitrary Prandtl number and viscosity. We discuss the results in Section 4. In Appendix A we delineate the
derivation of the dispersion equation for non-ideal plasma in the Boussinesq approximation 
for both adiabatic and non-adiabatic perturbations, and in Appendix B we find the analytical
solution of this dispersion equation for Keplerian discs at the neutral point.

\section{Linear analysis for ideal fluid}    

The dispersion relation for local small axially symmetric disturbances 
in the simplest case of an ideal fluid without entropy gradients 
reads (see \cite{1991ApJ...376..214B}, \cite{1998bhad.conf.....K} and Appendix A for the  derivation):
\beq{e:1}
\omega_*^4-\myfrac{k_z}{k}^2\kappa^2 \omega_*^2-4\Omega^2\myfrac{k_z}{k}^2 k_z^2 c_A^2=0\,.
\eeq
Here 
\beq{}
\omega_*^2=\omega^2-c_A^2k_z^2\,,
\eeq 
$k^2=k_r^2+k_z^2$, 
\beq{}
\kappa^2=4\Omega^2+r\frac{d\Omega^2}{dr}\equiv \frac{1}{r^3}\frac{d\Omega^2r^4}{dr}
\eeq
is the epicyclic frequency, and 
\beq{}
c_A^2=B_0^2/(4\piup\rho_0)
\eeq
in the unperturbed Alfv\'en velocity square. The initial magnetic field $B_0$ is assumed to be 
purely poloidal (directed along the $z$-coordinate) and homogeneous.  

The solution of the biquadratic equation (\ref{e:1}) has the form:
\beq{e:1sol}
\omega^2=\myfrac{k_z}{k}^2\left[c_A^2k^2+\frac{\kappa^2}{2}\pm \sqrt{\frac{\kappa^4}{4}+4\Omega^2c_A^2k^2}\right]\,.
\eeq
We will examine solutions of this equation by assuming $k_z^2/k^2\equiv k_z^2/(k_r^2+k_z^2)=const$, i.e. the direction of the wave vector in the $r-z$ plane is conserved; this is not
restrictive for our analysis. 
Depending on the sign of the root $\omega^2$, one of three modes can exist: 
the stable oscillating mode for $\omega^2>0$, indifferent 
equilibrium (neutral) mode for $\omega^2=0$, and exponentially growing mode for 
$\omega^2<0$.

According to the classical Rayleigh criterion \citep{LordRayleigh16}, if the epicyclic frequency $\kappa^2>0$
(in this case the angular momentum in the flow increases with radius), the equilibrium is stable.
If $\kappa^2<0$ (the angular momentum decreases with radius), the equilibrium is unstable.
If $\kappa^2=0$ (the angular momentum does not change with radius), the equilibrium is indifferent.   

\subsection{Ideal MHD case}


Let us start with discussing the behaviour of different modes of dispersion relation 
\eqn{e:1} in the ideal MHD case. It is instructive to investigate the asymptotics of 
these modes with decreasing (but non-zero) seed magnetic field
(see Section \ref{s:transition} for
more detail on the limiting transition for vanishing magnetic field).

If the magnetic field is present, there are five different types of solutions of \Eq{e:1sol} depending on how the angular velocity (angular momentum) changes with radius. 
\begin{figure}
\begin{center}
\centerline{\includegraphics[width=0.5\textwidth]{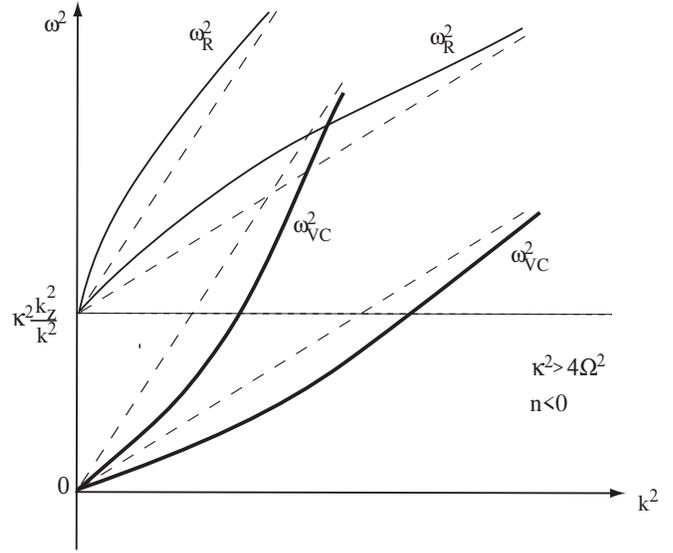}} 
\caption{Schematic behaviour of two branches of dispersion equation \Eq{e:1} ('Reynolds mode' $\omega_R^2$, thin curves,  and 'MRI mode' $\omega_{VC}^2$, thick curves) for 
two values of the Alfv\'en velocity $c_A^2$ (two values of the seed magnetic field $B_0$).
The dashed straight lines show the asymptotic behaviour of the solutions at large $k^2$: $\omega^2=(k_z/k)^2c_A^2k^2$. 
The smaller the seed magnetic field, the smaller the slope of the asymptotes.
Case 1 of the angular velocity and angular momentum increasing with radius ($\kappa^2>4\Omega^2$; $n<0$).}
\label{f:1}
\end{center}
\end{figure}
\begin{figure}
\begin{center}
\centerline{\includegraphics[width=0.5\textwidth]{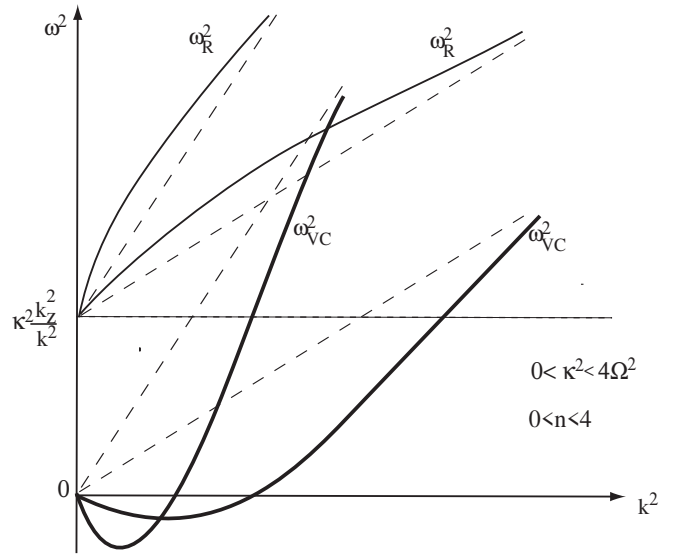}} 
\caption{The same as in Fig. 1 for the case of decreasing angular velocity with radius
but increasing angular momentum ($0<\kappa^2<4\Omega^2$; $0<n<4$) (case 2).}
\label{f:2}

\end{center}
\end{figure}
\begin{figure}
\begin{center}
\centerline{\includegraphics[width=0.5\textwidth]{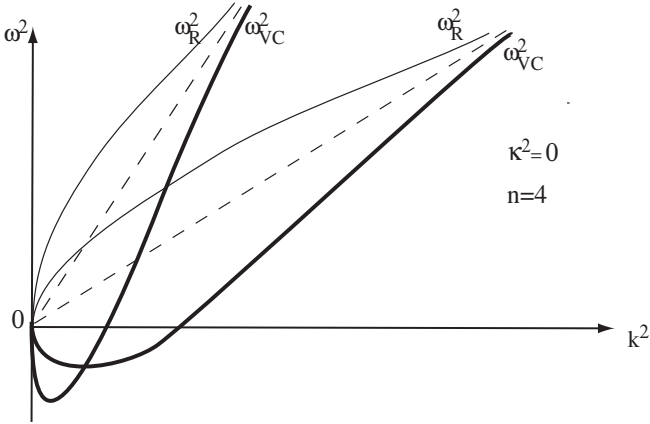}} 
\caption{The same as in Fig. 1 for the case of constant angular momentum ($\kappa^2=0$; $n=4$) (case 3).
Both the Rayleigh and MRI branches have infinite derivatives $d\omega^2/dk^2$ at $k^2=0$.}
\label{f:3}
\end{center}
\end{figure}

\textbf{Case 1: $\kappa^2>4\Omega^2$, $n<0$}. In this case there are two stable modes (see Fig. \ref{f:1}), which at large $k^2$ (short-wavelength limit) tend to the asymptotic behaviour $\omega^2=(k_z/k)^2c_A^2k^2$.
With decreasing (but non-zero) seed magnetic field amplitude $B_0$ (and the corresponding unperturbed Alfv\'en velocity $c_A$), one mode tends to the classical Rayleigh branch $\omega_R^2=(k_z/k)^2\kappa^2$ (the horizontal dashed line in Fig. 1),
and the second mode tends to the neutral branch $\omega_{VC}^2\to 0$.  

\textbf{Case 2: $0<\kappa^2<4\Omega^2$, $0<n<4$}. In this case the Rayleigh mode $\omega_R^2$ behaves almost in the same way as in case 1 (upper curves in Fig. \ref{f:2}). For the mode $\omega_{VC}^2$ (lower thick curves in Fig. \ref{f:2}) the instability arises in the interval: $0<k^2c_A^2<n\Omega^2$. It is in this case that the MRI instability occurs in a Keplerian accretion disc with $n=3$ and $\kappa=\Omega$. With decreasing $B_0$ the critical wave number separating the stable and unstable behaviour 
\beq{e:kcr}
k_{cr}^2(\omega^2=0)=n\frac{\Omega^2}{c_A^2}
\eeq
tends to infinity. The maximum instability growth rate 
characterized by the minimum of the mode $\omega^2_{VC}$ occurs at 
\beq{e:kmax}
k_{max}^2 =  \frac{n(8-n)}{16}\frac{\Omega^2}{c_A^2}\,.
\eeq
By substituting \Eq{e:kmax} into \Eq{e:1sol}, we find for the MRI mode
\beq{e:omega2max}
\omega^2_{VC,max}=-\frac{n^2}{16}\myfrac{k_z}{k}^2\Omega^2=-\frac{n}{8-n}\myfrac{k_z}{k}^2c_A^2k_{max}^2\,.
\eeq
With decreasing (but non-zero) $B_0$ and $c_A^2$, $\omega_{VC}^2(k_{max}^2)\to -0$  as $k_{max}^2\to \infty$.   

\textbf{Case 3: $\kappa^2=0$, $n=4$}. In this case (see Fig. \ref{f:3}) both the
Rayleigh mode $\omega_R^2$ and the MRI mode $\omega_{VC}^2$ go out of zero with infinite derivatives
(positive and negative for the Rayleigh and MRI modes, respectively).  With finite seed
magnetic field, the $\omega_{VC}^2$ mode displays the MRI. As $B_0$ becomes small (but non-zero), both modes asymptotically approach the neutral mode $\omega^2\to 0$.

\begin{figure}
\begin{center}
\centerline{\includegraphics[width=0.5\textwidth]{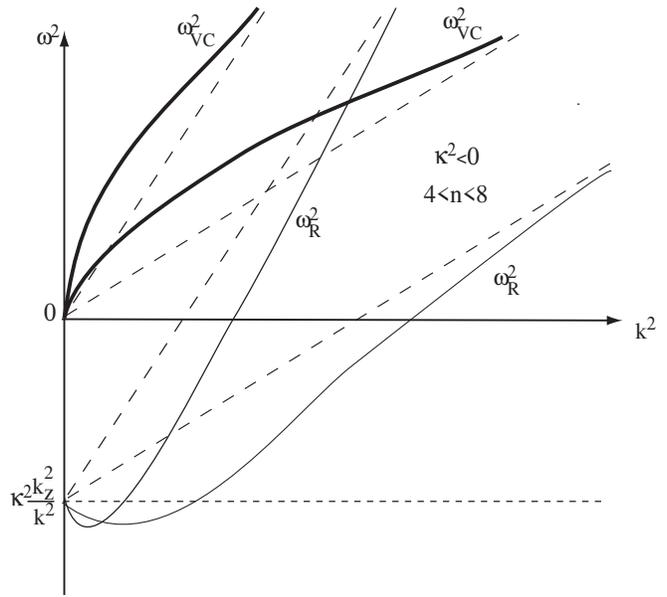}} 
\caption{The same as in Fig. 1 for the case of decreasing angular momentum ($\kappa^2<0$; $4<n<8$) 
(Case 4).
Instability according to the Rayleigh criterion occurs. The Rayleigh branch has a negative derivative at $k^2=0$.}
\label{f:4}
\end{center}
\end{figure}

\textbf{Case 4: $\kappa^2<0$, $4<n<8$}. In this case (see Fig. \ref{f:4}) in the absence of magnetic field the instability according to the Rayleigh criterion takes place (the bottom dashed horizontal line in Fig. 4) with $\omega_R^2=\kappa^2 (k_z/k)^2$. If the magnetic field is present, the Rayleigh instability is stabilized by the magnetic field at $k^2>k^2_{cr}$ (bottom thin curves in Fig. 4). Note that $k_{cr}^2$ and $k_{max}^2$ 
here are the same as in Case 2. While similar to the MRI mode, this is now \textit{the Rayleigh mode} 
$\omega_R^2$ that is unstable and reaches maximum growth rate $\omega_{R,max}^2$ determined by 
\Eq{e:omega2max}. 
In contrast, \textit{the Velikhov-Chandrasekhar mode} $\omega_{VC}^2$ (upper thick curves in Fig. 4) remains stable \textit{at all wavenumbers}, and with decreasing (but non-zero) 
magnetic field $\omega_{VC}^2\to +0$. We stress again that 
the difference between the Rayleigh and MRI modes is due to their different asymptotic behaviour 
as $B_0\to +0$: the Rayleigh mode is unstable and 
behaves as $\omega_R\to -\kappa^2 k_z^2/k^2$, unlike the stable Velikhov-Chandrasekhar mode. 

\begin{figure}
\begin{center}
\centerline{\includegraphics[width=0.5\textwidth]{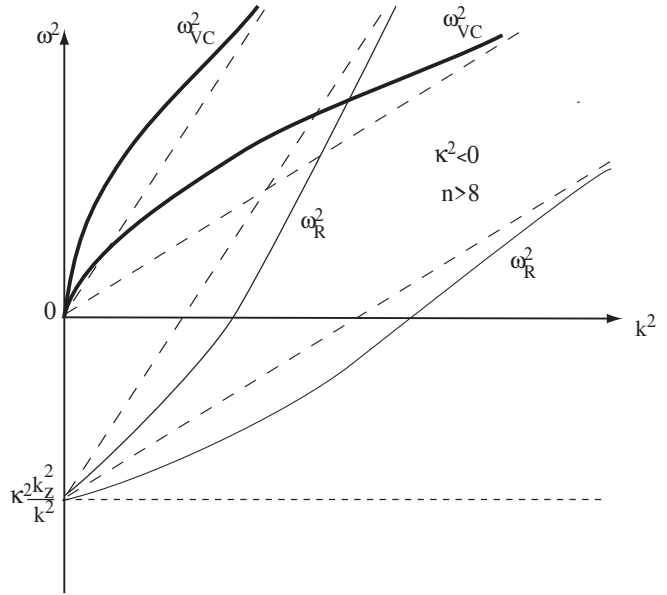}} 
\caption{The same as in Fig. 4 for the case ($\kappa^2<0$; $n>8$) (case 5 in the text); the Rayleigh branch has a positive
derivative at $k^2=0$.}
\label{f:5}
\end{center}
\end{figure}

\textbf{Case 5: $\kappa^2<0$, $n>8$}. The only difference of this case from Case 4 is that the Rayleigh mode $\omega_R^2$ goes out of zero with a positive derivative (bottom thin curves in Fig. \ref{f:5}).


\subsection{On the behaviour of MRI at vanishing magnetic field}
\label{s:transition}

The transition to purely
hydrodynamic case without magnetic field should be treated separately. 
Let us consider asymptotic 
solutions \eqn{e:1sol} for vanishing magnetic field. In the leading order in $c_A$ two 
branches of the dispersion relation are:
\beq{omegaR}
\omega^2_{R}\simeq \myfrac{k_z}{k}^2\left[\kappa^2+
c_A^2k^2\left(1+\frac{4\Omega^2}{\kappa^2}\right)\right]\,,
\eeq
which we have refered to as the Rayleigh mode 
since in the absence of the magnetic field it 
tends to the classical Rayleigh mode $\omega^2_R=(k_z/k)^2\kappa^2$, and
\beq{omegaVC}
\omega^2_{VC}\simeq {k_z}^2c_A^2\left(1-\frac{4\Omega^2}{\kappa^2}\right)\,,
\eeq
which we have refered to as the Velikhov-Chandrasekhar mode and which is 
manifestly unstable for the Keplerian motion ($\kappa^2=\Omega^2$). 

Notice that unlike the Rayleigh mode, setting magnetic field to zero in \Eq{omegaVC} 
leads to a paradoxical result: $\omega^2_{VC}=0$. This 'neutral mode' is fictitious, it
does not exist in the purely hydrodynamic case. To see this, let us write linearized 
system of perfect fluid equations in the Boussinesq approximation 
(see \eqn{Bq} -- \eqn{iuz} and \Eq{rho1} in Appendix A):
\begin{eqnarray}
&k_ru_r+k_zu_z=0\nonumber\\
&i\omega u_r-2\Omega u_\phi=ik_r\frac{p_1}{\rho_0}\nonumber\\
&i\omega u_\phi+\frac{\kappa^2}{2\Omega}u_r=0\nonumber\\
&i\omega u_z=ik_z\frac{p_1}{\rho_0}
\label{ideal}
\end{eqnarray}
It is easy to find the dispersion relation in this case:
\beq{iddisp}
\omega^2=\myfrac{k_z}{k}^2\kappa^2\,,
\eeq
which is the classical Rayleigh branch. No neutral mode $\omega^2=0$ arises.
The neutral mode $\omega=0$ does exist in the purely
hydrodynamic case but only for specific choice of radial perturbations
with $u_r=u_z=k_z=0$ and $-2\Omega u_\phi=ik_r(p_1/\rho_0)$ (see \eqn{ideal}).
The odd mode $\omega^2=0$ arising in the limiting transition with vanishing magnetic field formally
appears from \Eq{e:1} because the fourth order of this dispersion relation is entirely due to 
the square brackets $\sim \omega^2$ in the denominator of \Eq{uphi}, which in the case $B=0$ cancels
with the brackets $\sim \omega^2$ in the nominator.

Similarly, no  
smooth transition to the hydrodynamic case occurs if viscosity is included (see below).  
The absence of the smooth transition to the ideal hydrodynamic case when $B\to 0$ was first 
noted by Velikhov (1959). At the same time, the transition to the classical Rayleigh mode 
with vanishing magnetic field occurs smoothly.

\section{Linear analysis for fluid with viscosity and magnetic diffusivity}

Consider the more general case of a non-ideal viscous fluid 
with finite electric conductivity characterized by the kinematic 
viscosity coefficient $\nu$
 and resistivity (magnetic diffusivity) $\eta$. Naturally, in problems with viscosity
and magnetic diffusivity there is no initial steady state. The angular momentum is redistributed
by viscosity 
on the time scale $\tau_\nu\sim R^2/\nu$, and the magnetic field changes on
the magnetic diffusion time scale $\tau_\eta\sim R^2/\eta$, where $R$ is the characteristic size 
of the system. Everywhere below we will assume these timescales to be 
extremely long compared to the Keplerian rotation time and the characteristic
instability growth time, if conditions are suitable for the latter to arise.  
Dispersion relation in this case 
can be derived following the local linear analysis of MRI performed, e.g., in 
the monograph \cite{1998bhad.conf.....K} with taking into account viscosity and 
conductivity in the WKB-approximation (see Appendix A, with zero density perturbations
\Eq{rho1}):
\beq{e:eq*}
\omega_{**}^4+\myfrac{k_z}{k}^2\left[\left(i\omega+\eta k^2\right)^2\kappa^2+c_A^2k_z^2
(\kappa^2-4\Omega^2)\right]=0\,,
\eeq
where
\beq{e:omega**}
\omega_{**}^2=-(i\omega + \nu k^2)(i\omega + \eta k^2)-c_A^2k_z^2\,.
\eeq
The dispersion relation \eqn{e:eq*} is identical to the one derived for a
rotating liquid metal
annulus in the incompressible limit  \citep{2001MNRAS.325L...1J}\footnote{Note that 
those authors searched for a stable differential rotation law between cylinders
with given viscosity and electric conductivity while we are investigating 
conditions for MRI in a viscous, electrically conducting flow in gravitational field with 
given differential rotation law.}. This equation was also derived and mathematically analysed in
\cite{2008ApJ...684..498P}. However, that paper focused on the application of the 
MRI mode to the calculations of the Reynolds and Maxwell stresses in the differentially rotating
flow. In what follows we shall discuss the constraints on MRI modes in astrophysical accretion discs,
where the free-path length of particles (and hence the viscosity) 
is limited by the disc thickness.

The magnetic Prandtl number is introduced as $\mathrm{P_m}=\nu/\eta$. Using the standard 
expressions for $\nu$ and $\eta$ for fully ionized hydrogen plasma \cite{1962pfig.book.....S}, 
we readily find 
\beq{e:Pm}
\mathrm{P_m}\approx 3.4\times 10^{-28}\frac{T^4}{\rho\ln \Lambda_{eH}\Lambda_{pH}}\,,
\eeq
where $T$ is the temperature, $\rho$ is the density and $\Lambda_{eH}$ and $\Lambda_{pH}$ are
electron and proton Coulomb logarithms, respectively.

As was shown
by \cite{2008ApJ...674..408B}, the magnetic Prandtl number can be of the order
of one in the inner parts 
of accretion discs around neutron stars and black holes.

\subsection{The case of the magnetic Prandtl  number  $\mathrm{P_m}$=1}  

Here we will discuss the exact analytic solution to \Eq{e:eq*}
for the important particular case $\mathrm{P_m}= 1$ (which can be derived,
for example, from the general analytic solution found in \cite{2008ApJ...684..498P}) and obtain restrictions on 
the maximum mean free-path length of ions in accretion discs at which 
MRI disappears due to non-ideality effects. 

The exact solution of \Eq{e:eq*} for $\mathrm{P_m}= 1$ is
\beq{e:sol*}
\omega=i\nu k^2\pm \sqrt{
\myfrac{k_z}{k}^2
\left[c_A^2k^2+\frac{\kappa^2}{2}\pm 
\sqrt{\frac{\kappa^4}{4}+4\Omega^2c_A^2k^2}
\right]
}\,.
\eeq
Here the plus sign before the second square root corresponds to the Rayleigh branch, and the 
minus sign corresponds to the Velikhov-Chandrasekhar (MRI) branch. We shall examine 
below the MRI branch only. 

It is noted that the first square root in this equation 
contains the solutions \eqn{e:1sol} of \Eq{e:1}:
\beq{}
\omega=i\left(\nu k^2 - \sqrt{-\omega^2_{\nu=0}}\right)\,.
\eeq
(Here we remind that for regions with  MRI $\omega^2<0$). Also note that like 
in the ideal MHD case considered above in Section \ref{s:transition}, 
here there are no smooth transition of 
the MRI mode to the hydrodynamic case with viscosity when vanishing the
magnetic field. As can be straightforwardly derived from \Eq{Bq}-\eqn{iuz} in 
the Appendix, the dispersion relation for the hydrodynamic case with viscosity reads:
\beq{visc}
(i\omega+\nu k^2)^2+\myfrac{k_z}{k}^2\kappa^2=0\,.
\eeq 
While the Rayleigh mode (the one with positive sign before the second square root in 
\Eq{e:sol*}) tends to the mode given by \Eq{visc} when magnetic field is vanishing, 
the MRI mode (the one with positive sign before the second square root in 
\Eq{e:sol*}) completely disappears (there is no mode $i\omega +\nu k^2=0$ 
without magnetic field, unless $k_z=0$).  

Below we will consider the case $k_z=k$, i.e. with $k_r=0$. 
For further analysis it is convenient to 
rewrite the dispersion relation \eqn{e:eq*} in the dimensionless form.
We introduce the dimensionless variables: 
\beq{e:dimless}
\tilde \omega\equiv \omega/\Omega; 
\quad \tilde k\equiv \frac{c_A k}{\Omega}; \quad \tilde \kappa^2\equiv \kappa^2/\Omega^2; 
\quad \tilde \nu \equiv \frac{\nu \Omega}{c_A^2}
\eeq 
For Keplerian discs the dimensionless epicyclic frequency is $\tilde \kappa^2=1$. 
In the dimensionless variables, solution to \Eq{e:eq*} takes the form:
\beq{e:eqd}
\tilde\omega = i\left(\tilde\nu \tilde k^2\pm 
\sqrt{-\tilde k^2-\frac{1}{2}\mp \sqrt{\frac{1}{4}+4\tilde k^2}} \right)\,.
\eeq
Of the four solutions of \Eq{e:eqd} we choose the one for the MRI mode:
\beq{e:eqd1}
\tilde\omega = i\left(\tilde\nu \tilde k^2 -
\sqrt{-\tilde k^2-\frac{1}{2} + \sqrt{\frac{1}{4}+4\tilde k^2}} \right)\,.
\eeq
Now we find the neutral point $\tilde\omega =0$. Squaring twice 
\Eq{e:eqd1}, we obtain the equation for the critical wavenumber $\tilde k_{cr}$ separating 
unstable ($\tilde k<\tilde k_{cr}$) and stable ($\tilde k>\tilde k_{cr}$) perturbations:
\beq{e:kcrPm1}
\tilde \nu^4 \tilde k^6+2\tilde\nu^2 \tilde k^4+(1+\tilde \nu^2) \tilde k^2 -3 =0\,.
\eeq
Without viscosity we recover the old result: $\tilde k_{cr}^2=3$ 
(see \Eq{e:kcr}). It is easy to check that for the dimensionless viscosity  $\tilde\nu=4/5$ the 
neutral point is $\tilde k_{cr}=\sqrt{15/16}$, i.e. here the neutral 
point coincides with the maximum wavenumber $k_{max}$ at which the maximum MRI growth occurs
in the inviscid case (see \Eq{e:kmax} above).   
At large dimensionless viscosity $\tilde \nu\gg 1$, the asymptotic solution of \Eq{e:eqd1} reads
\beq{e:kcras}
\tilde k_{cr}\simeq \frac{\sqrt{3}}{\tilde \nu}\,.
\eeq 
Therefore, at arbitrarily high viscosity there exists the interval of wavenumbers 
$0<\tilde k<\tilde k_{cr}$
where MRI is still takes place, but the MRI increment here is very small.

\begin{figure}
\begin{center}
\centerline{\includegraphics[width=0.5\textwidth]{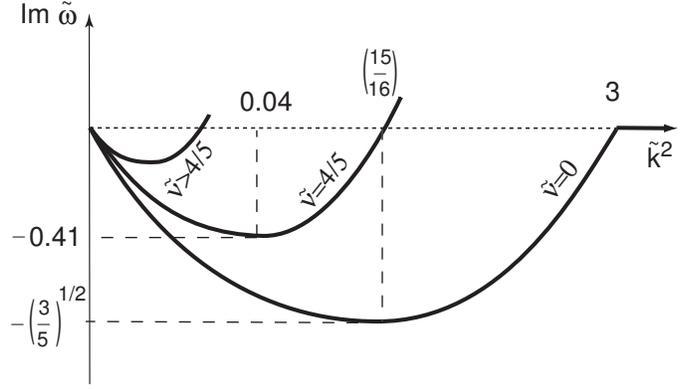}} 
\caption{Schematics of the influence of viscosity on the MRI condition $0<\tilde k<\tilde k_{cr}$.
Shown are curves of the imaginary part of $\tilde\omega$ as a function of the dimensionless
wave number $\tilde k^2$.
With increasing viscosity, the MRI interval shifts to the left and shrink (see also Fig. 1 in Pessah \& Chen (2008)).}
\label{f:mri}
\end{center}
\end{figure}

Actually, in realistic accretion discs with finite thickness $H$
we should take into account that there is the lower limit for $k$ corresponding
to the obvious restriction on the maximum perturbation wavelength $\lambda<2H$:
\beq{}
k = \frac{2\piup}{\lambda}>\frac{\piup}{H}\equiv k_{min}\,.
\eeq
Therefore, in the dimensionless variables we find the MRI condition in the form:
\beq{e:MRI1}
\tilde k_{min}\le \tilde k\le \tilde k_{cr}\,.
\eeq
It is also convenient to change from the disc thickness $H$ to the
characteristic thermal velocity in the disc $c_s$, since in accretion discs  
the hydrostatic equilibrium along the vertical coordinate yields 
\beq{e:cs}
c_s=\Pi \Omega H
\eeq
where $\Pi$ is a numerical coefficient.
For example,  in the standard geometrically thin Shakura-Sunyaev 
$\alpha$-disk $\Pi=1/\sqrt{4\Pi_1}\simeq 1/\sqrt{20}$
(see \cite{1998A&AT...15..193K}). Then in the inviscid fluid $\tilde k_{cr}=\sqrt{3}$, 
$\tilde k_{min}=\piup \Pi (c_A/c_s)$, and the MRI condition \Eq{e:MRI1} takes the form
\beq{e:MRI2}
\piup \Pi \myfrac{c_A}{c_s}\le \sqrt{3}\,.
\eeq
Essentially, this is the well-known condition that for MRI to operate the 
seed magnetic field should not exceed some critical value.

\begin{figure*}
\begin{center}
\centerline{\includegraphics[width=0.8\textwidth]{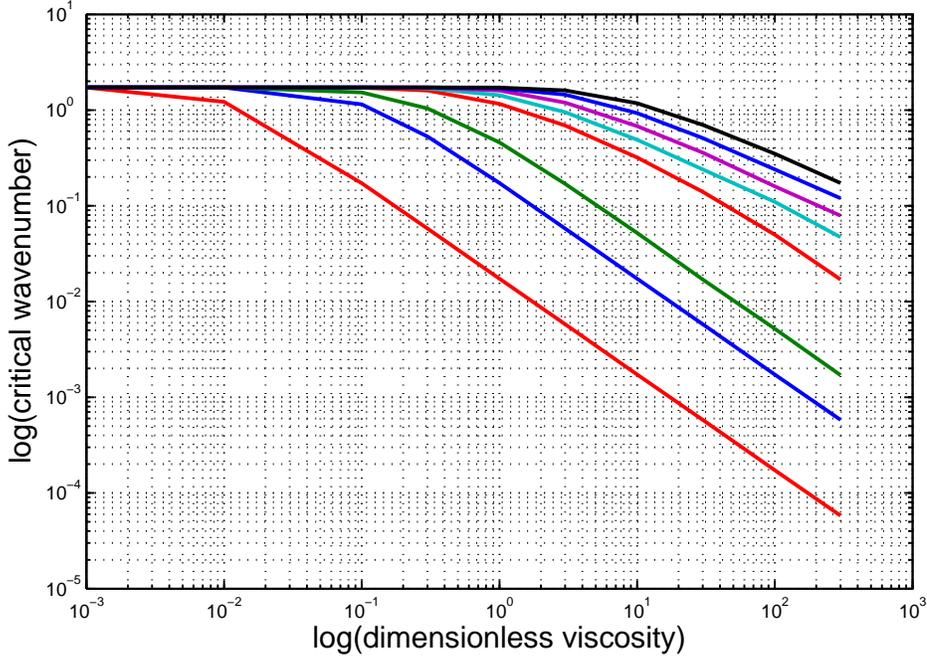}} 
\caption{Dimensionless critical wavenumber $\tilde k_{cr}$ as a function of
dimensionless viscosity coefficient $\tilde \nu$ for different magnetic
Prandtl numbers  $\mathrm{P_m}$. Lines from bottom to top correspond to  $\mathrm{P_m}$=0.01, 0.1,
0.3, 3, 10, 30, 100, 300.}
\label{f:mripl}
\end{center}
\end{figure*}

In the non-ideal plasma the MRI condition \Eq{e:MRI2} becomes
\beq{e:MRI3}
\piup\Pi\myfrac{c_A}{c_s}\le \tilde k_{cr}\,.
\eeq
Note that $\tilde k_{cr}$ decreases with $\tilde \nu$. For example, 
if $\tilde \nu$ is high, \Eq{e:kcras} implies very small values of $\tilde k_{cr}$
and, correspondingly, very low $c_A$ at which MRI can occur with 
uninterestingly small increments. The schematic behaviour of the MRI mode 
at non-zero viscosity is shown in Fig. \ref{f:mri}. At arbitrary finite viscosity 
$\tilde\nu$ the neutral point $\tilde\omega(\tilde k_{cr})$ separates 
exponentially growing small perturbations $\propto exp(i\omega t)$ (the lower part of Fig. \ref{f:mri}
where Im$\tilde \omega>0$) from 
exponentially decaying ones (the upper part of Fig. \ref{f:mri}). At zero viscosity,
however,  the function $\tilde\omega (\tilde k)$ (the curve labeled by $\tilde\nu=0$)
ends at the point $\tilde k_{cr}=\sqrt{3}$, because in this case at $\tilde k\ge k_{cr}$ 
the $\tilde\omega$ becomes purely real and small perturbations oscillate.

In the case of high viscosity it is convenient 
to express the ratio $c_A/c_s$ through the dimensionless viscosity 
$\tilde \nu$. Using the conventional definition of the viscosity coefficient
$\nu=c_sl$, where  $l$ is the effective mean-free path of ions with account for the  
Coulomb logarithm, and our convention for the thermal velocity in the disc \eqn{e:cs} introduced above, we find:
\beq{e:tnu}
\tilde \nu \equiv \nu \frac{\Omega}{c_A^2}=\frac{1}{\Pi} \myfrac{c_s}{c_A}^2\myfrac{l}{H}\,.  
\eeq
Finally, we obtain the MRI condition in the convenient form:
\beq{e:MRIf}
\frac{l}{H}\le \frac{1}{\piup^2 \Pi}\tilde\nu \tilde k_{cr}^2\,.
\eeq

In the particular case  $\mathrm{P_m}$=1 we can explicitly find $\tilde \nu \tilde k_{cr}^2$ from 
\Eq{e:eqd1}:
\beq{e:nuk2}
\tilde\nu \tilde k^2_{cr} =
\sqrt{-\tilde k^2_{cr}-\frac{1}{2}+ \sqrt{\frac{1}{4}+4\tilde k^2_{cr}}}\,, 
\eeq
so that condition \eqn{e:MRIf} takes the form:
\beq{e:MRIfP1}
\frac{l}{H}\le \frac{1}{\piup^2 \Pi}\sqrt{-\tilde k_{cr}^2-\frac{1}{2}+\sqrt{\frac{1}{4}+4\tilde k_{cr}^2}}\,.
\eeq
(This formula should be used when $\nu\ne 0$, i.e. when $\tilde k_{cr}^2<3$).
Consider first the case of small viscosities where $\tilde k_{cr}^2\approx 3$.
By introducing the small parameter $\epsilon=3-\tilde k_{cr}^2\ll 1$ and expanding 
the left-hand side of \Eq{e:MRIfP1} in $\epsilon$, we obtain
\beq{e:MRIy}
\frac{l}{H}\le \frac{1}{\piup^2 \Pi}\sqrt{\frac{41}{49}\epsilon}\,.
\eeq

Now consider the special case where $\tilde k_{cr}$ coincides with the wavenumber of maximum MRI increment in the ideal fluid:  
$\tilde k_{cr}=\tilde k_{max}=\sqrt{\frac{15}{16}}$
(see \Eq{e:kmax}). This is realized at $\tilde \nu=4/5$. Here we find the limit
\beq{}
\myfrac{l}{H}\le \frac{1}{\piup^2 \Pi} 0.75\approx 0.34.
\eeq

Finally, in the high-viscosity limit for  $\mathrm{P_m}$=1 $\tilde \nu\gg 1$,
substituting the asymptotic \eqn{e:kcras} into \Eq{e:MRIf} with account for
the expression for dimensionless viscosity \eqn{e:tnu} we obtain 
\beq{e:MRIhightnu}
\myfrac{l}{H}\le \frac{\sqrt{3}}{\piup}\myfrac{c_A}{c_s}\,,\quad \mathrm{P_m}=1, \tilde\nu\gg 1\,.
\eeq
Note that this constraint is insensitive to the disc vertical structure
parameter $\Pi$.
This condition can be checked for particular microphysics plasma
properties in different thin Keplerian discs.

\subsection{The case of arbitrary magnetic Prandtl number}


The generalization of the above analysis to an arbitrary Prandtl number is
straightforward. First, for given  $\mathrm{P_m}$ and $\tilde \nu$ we solve the dimensionless \Eq{e:eq*} to
find $\tilde k_{cr}(\tilde\nu, \mathrm{P_m})$ (see Appendix \ref{A:analyt}), and then obtain the general MRI condition \eqn{e:MRI3}
\beq{}
\myfrac{c_A}{c_s}\le \frac{1}{\piup \Pi} \tilde k_{cr}(\tilde\nu, \mathrm{P_m})\,.
\eeq

The result of calculation of $\tilde k_{cr}$ for a range of magnetic Prandtl
numbers $\mathrm{P_m}$ and dimensionless viscosities $\tilde\nu$ can be 
found in \cite{2008ApJ...684..498P} (see e.g. their Fig. 6 and 7) and 
is illustrated in Fig. \ref{f:mripl}.

In the limiting case of high dimensionless viscosities $\mathrm{P_m}/\tilde \nu^2 \ll 1$, which 
can be realized in the outer parts of thin Keplerian accretion discs (see \Eq{e:Pm} above),
using asymptotic \eqn{e:kcrasa} and definition \eqn{e:tnu}, we find the
restriction on the mean-free path of ions in the disc 
\beq{e:MRIPsmall}
\myfrac{l}{H}\le \frac{\sqrt{3}\mathrm{P_m}}{\piup}\myfrac{c_A}{c_s}\,,
\quad \mathrm{P_m}/\tilde\nu^2 \ll 1\,.
\eeq
which is the generalization of \Eq{e:MRIhightnu} for arbitrary magnetic Prandtl number.
Using the expression for the dimensionless viscosity \eqn{e:tnu}, the 
condition for the power-law asymptotic $\mathrm{P_m}/\tilde\nu^2 \ll 1$ can be
recast to the inequality 
\beq{}
\mathrm{P_m}/\tilde\nu^2 \ll 1\, \Leftrightarrow 
\myfrac{l}{H}^2\gg \Pi \mathrm{P_m}\myfrac{c_A}{c_s}^4\,.
\eeq
Therefore, the MRI condition can be written in terms of the interval for 
$l/H$ in a Keplerian disc as 
\beq{e:MRIPnu}
\sqrt{\Pi\mathrm{P_m}}\myfrac{c_A}{c_s}^2\ll \myfrac{l}{H}\le \frac{\sqrt{3}\mathrm{P_m}}{\piup}\myfrac{c_A}{c_s}\,.
\eeq

\section{Discussion and conclusion}

In the present paper we have extended the original analysis of MRI in
ideal MHD plasmas carried out by
\cite{2012MNRAS.423L..50B}.  
First, we emphasize that 
hydromagnetic flows in which the angular momentum increases or decreases with radius are different 
from the point of view of the MRI development. 
In the classical Rayleigh-unstable case  where the angular momentum 
decreases with radius, the Velikhov-Chandrasekhar MRI mode is stable, while
the Rayleigh mode is unstable (see Fig 4, 5); 
the magnetic field stabilizes the Rayleigh mode in the 
short-wavelength limit. When the angular momentum in the flow 
increases with radius, MRI arises  
at long wavelengths (small wave numbers $k$, see Fig. 2). 
However, the local WKB approximation should be applied with caution 
at long wavelengths. At long wavelengths, 
the ansatz for the solution should be rather taken 
in the global form $f(r)e^{i(\omega t-k_rr-k_zz)}$. Note that the original papers by Velikhov and Chandrasekhar analyzed the linear stability of magnetized flows between cylinders exactly in that approximation (see also \cite{1999ApJ...515..776S} for the global analysis of perturbations in an inviscid magnetized proto-planetary discs 
with non-zero magnetic diffusivity). 

Second, in the phenomenologically interesting case of thin Keplerian accretion discs, viscosity may restrict MRI growth. This situation can be realized in the inner parts of accretion discs. Indeed, at high temperatures the mean free path 
of ions $l\sim T^2$ can become comparable with the characteristic disc thickness $H$
at $H<r$ (thin discs). This means that the flow should be treated 
kinetically (see, for example, recent 2.5D hybrid calculations \cite{2014PhPl...21e2903S}
or the discussion of MRI in rarefied astrophysical plasmas with Braginskii viscosity 
in \cite{2005ApJ...633..328I}). 
The seed small magnetic field under these conditions does 
not grow, i.e. the high ion viscosity can suppress MRI. 
Clearly, this interesting regime requires further study. 

At large magnetic Prandtl numbers $\mathrm P_m\gg 1$, which can be realized in the innermost parts of accretion discs around neutron stars and black holes, 
the kinematic viscosity $\nu$ is much larger than the magnetic diffusivity $\eta$. In this case 
plasma may become collisionless, and hydrodynamic description fails. Our analysis
shows that in principle the collisionless regime (the ion mean-free path comparable to or larger 
than the disc thickness, $l\sim H$) in 
Keplerian discs 
can be realized even for magnetic Prandtl numbers $\mathrm P_m\simeq 1$
(see \Eq{e:MRIPnu}).

We have also obtained the dispersion relation for local small perturbations in the Boussinesq limit for non-adiabatic perturbations
(see \Eq{e:eq*gen}). This is the fifth-order algebraic equation, in contrast to the fourth-order dispersion relation for adiabatic perturbations or non-adiabatic perturbations with $k_r=0$ in non-ideal plasma \eqn{e:eq*}. Also note that when the density perturbations are expressed through the entropy gradients (see Eq. (2.2h) in \cite{1991ApJ...376..214B}), the frequency appears in the denominator but the final dispersion relation (2.5) in \cite{1991ApJ...376..214B}  
remains to be the fourth-order equation in $\omega$ even with taking into account the entropy gradients. 
Apparently, the difference is due to the fact that in the case of non-adiabatic perturbations the density variations are proportional to the azimuthal velocity perturbations $u_\phi$ (see our \Eq{e:rho1uphi}) and not to 
$u_z$ and $u_r$ as in the case considered by \cite{1991ApJ...376..214B}. 
The analysis of the effect of non-adiabatic perturbations deserves 
a separate study and will be addressed in a future work.

Perturbations with $k_r=0$ represent waves propagating along the $z$-coordinates, and when their wavelength is comparable to the disc thickness, the WKB approximation becomes problematic. Perturbations with $k_z=0$ propagate along the $r$-coordinate, which is much larger than the disc thickness for thin accretion discs. However, for such perturbations with $k=k_r$ and $k_z=0$ the second term in \Eq{e:eq*} and \Eq{e:eq*gen} vanishes, and therefore from \Eq{e:omega**} we find two perturbation modes 
\beq{}
\omega_1=i\nu k^2\,,\quad \omega_2=i\eta k^2\,,
\eeq   
i.e. decaying standing waves for any seed magnetic field. This may suggest that in the poloidal magnetic fields purely radial perturbations with $k=k_r$ do not grow. The situation is different when the azimuthal magnetic field is present. This case should be considered separately and has been investigated for a range of astrophysical applications in other papers (see, e.g., \cite{1978RSPTA.289..459A, 1999ApJ...515..776S, 2014arXiv1407.0240R, 2014JFM...760..591K}).

We conclude that in thin Keplerian 
accretion discs the adding of viscosity 
can strongly restrict the MRI conditions once the mean free path of ions becomes comparable with 
the disc thickness. These limitation should be taken into account in the 
direct numerical simulations of MRI in astrophysical accretion discs.  

\section{Acknowledgements}

We thank the anonymous referee for drawing our attention to 
earlier papers by \cite{2008ApJ...684..498P}, \cite{2005ApJ...633..328I} 
and \cite{2008ApJ...689.1234M} and for the constructive criticism.
We also thank Prof. Dr. F. Meyer for discussions and MPA (Garching) for hospitality. 
The work was supported by the Russian Science Foundation grant 14-12-00146.

\appendix
\section{Derivation of the dispersion equation for non-ideal
plasma}

Here we generalize the derivation of the MRI dispersion equation \eqn{e:1}
given in \cite{1998bhad.conf.....K} for the case of non-ideal plasma with
arbitrary kinetic coefficients $\nu$ and $\eta$ (see also \cite{2001MNRAS.325L...1J}). 

The system of non-deal MHD equations reads:

1) mass conservation equation
\beq{}
\frac{\partial \rho}{\partial t}+\nabla\cdot(\rho\bm{u})=0\,,
\eeq
2) Navier-Stokes equation including gravity force and Lorentz force
\beq{}
\frac{\partial \bm{u}}{\partial t}+(\bm{u}\nabla)\cdot\bm{u}=-\frac{1}{\rho}\nabla p -
\nabla \phi_g +
\frac{1}{4\piup\rho}(\nabla\times \bm{B})\times \bm{B}+\nu\Delta\bm{u}
\eeq
(here $\phi_g$ is the Newtonian gravitational potential), 

\noindent 3) induction equation
\beq{}
\frac{\partial \bm{B}}{\partial t}=\nabla\times (\bm{u}\times \bm{B})+\eta\Delta\bm{B}\,,
\eeq
4) energy equation 
\beq{e:en}
\frac{\rho {\cal R} T}{\mu}\left[
\frac{\partial s}{\partial t}+(\bm{u}\nabla)\cdot s
\right]=
Q_{visc}-\nabla\cdot\bm{F}+\frac{\eta}{4\piup}[\nabla\times \bm{B}]^2\,.
\eeq
where $s$ is the specific entropy (per particle), ${\cal R}$ is the universal gas constant, 
$\mu$ is the molecular weight, $T$ is the temperature, and terms on the right 
stand for viscous, energy flux $\bm{F}$ and Joule dissipation, respectively. 

5) These equations should be completed with the equation of state for a perfect gas, which is convenient to write in the form:
\beq{e:eos}
p=Ke^{s/c_V}\rho^\gamma\,,
\eeq 
where $K$ is a constant, $c_V$ is the specific volume heat capacity 
and $\gamma=c_p/c_V$ is the adiabatic index (5/3 for the monoatomic gas). 

We will consider small axially symmetric perturbations in 
the WKB approximation with space-time dependence $e^{i(\omega t-k_r r-k_z z)}$, where 
$r,z,\phi$ are cylindrical coordinates. The unperturbed 
magnetic field is assumed to be purely poloidal: $\bm{B_0}=(0, 0, B_0)$. 
The velocity and magnetic field perturbations are $\bm{u}=(u_r,u_\phi,u_z)$ and 
$\bm{b}=(b_r, b_\phi, b_z)$, respectively. 
The density, pressure and entropy perturbations are $\rho_1$, $p_1$, and $s_1$ over the unperturbed values $\rho_0$, $p_0$, and $s_0$, respectively. To filter out magnetoacoustic oscillations arising from the restoring pressure force, we
will use the Boussinesq approximation, i.e. consider incompressible gas motion 
$\nabla\cdot \bm{u}=0$. 
In the energy equation we neglect Eulerian pressure variations,
$p_1(t, r, \phi, z) =0$, but Lagrangian pressure variations
$\delta p(t, r(t_0), \phi(t_0, z(t_0))$ are non-zero. (We remind that for infinitesimally 
small shifts the perturbed gas parcel acquires the pressure equal to that of the ambient medium; 
see e.g. \cite{1960ApJ...131..442S,  Kundu5ed} for discussion of the Boussinesq approximation). 

In the linear approximation, the system of differential non-ideal MHD equations is reduced to the following system of algebraic equations.

a) The Boussinesq approximation for gas velocity $\bm{u}$ is $\nabla\cdot  \bm{u}=0$:
\beq{Bq}
k_r u_r+k_zu_z=0\,.
\eeq

b) The radial, azimuthal and vertical components of the Euler momentum equation are, respectively:
\beq{iur}
i\omega u_r-2\Omega u_\phi=ik_r\frac{p_1}{\rho_0}-\frac{\rho_1}{\rho_0^2}\frac{\partial
p_0}{\partial r}+i\frac{c_A^2}{B_0}(k_rb_z-k_zb_r)-\nu k^2u_r\,,
\eeq
\beq{iuphi}
i\omega u_\phi+\frac{\kappa^2}{2\Omega}u_r=-i\frac{c_A^2}{B_0}k_zb_\phi-\nu k^2u_\phi\,,
\eeq
\beq{iuz}
i\omega u_z=ik_z\frac{p_1}{\rho_0}-\frac{\rho_1}{\rho_0^2}\frac{\partial p_0}{\partial z}-\nu k^2u_z
\eeq
Here $k^2=k_r^2+k_z^2$ so that in the linear order $\nu \Delta \bm{u}\to -\nu k^2\{ u_r,u_\phi,u_z\}$\footnote{Here we neglect terms
$\sim(k_r/r)$ compared to terms $\sim k^2$, see also discussion in \cite{1978RSPTA.289..459A}.},
and we have introduced the unperturbed Alfv\'en velocity $c_A^2=B_0^2/(4\piup \rho_0)$.

To specify density perturbations $\rho_1/\rho_0$, we need to address the energy equation. 
First, consider adiabatic perturbations, i.e. require 
\beq{adiab}
\frac{\partial s}{\partial t} + (\bm{u}\cdot\nabla) s=0\,.
\eeq
For small density perturbations from \Eq{e:eos} we obtain for entropy perturbations
\beq{e:s1}
\frac{s_1}{c_V}+\gamma\frac{\rho_1}{\rho_0}=0\,,
\eeq
and after substituting this into \Eq{adiab} we get 
\beq{rho1ad}
 i\omega\gamma\frac{\rho_1}{\rho_0}+u_z\frac{\partial \ln p\rho^{-\gamma}}{\partial z} + u_r\frac{\partial \ln
p\rho^{-\gamma}}{\partial r}=0
\eeq
(cf. Eq. (122) in Balbus \& Hawley (1998)). Hence in the absence of entropy gradients we obtain 
\beq{rho1}
\frac{1}{\rho_0}\frac{\partial \rho_1}{\partial t}=0\,.
\eeq

Consider now the more general case of \textit{non-adiabatic} linear perturbations. 
To do this, we need to specify the right-hand side of the energy equation \eqn{e:en}. 
Let us start with the last term. Writing for the magnetic field $\bm{B}=\bm{B_0}+\bm{b}$ 
and taking into account that for the unperturbed field $\nabla\times \bm{B_0}=0$, 
we see that the Joule dissipation term is quadratic in magnetic field perturbations $\bm{b}$, 
so we exclude it from consideration. 
The heat flux divergence is 
\beq{e:kappat}
\nabla\cdot\bm{F}=\nabla(-\kappa_T\nabla T)=-\kappa_T\Delta T\,,
\eeq
where $\kappa_T$ is the temperature conductivity coefficient.
From equation of state for ideal gas written in the form 
$p=\rho {\cal R} T/\mu$, we find for small perturbations with zero Eulerian pressure variations 
$p_1/p_0=0$ 
\beq{e:T1}
\frac{\rho_1}{\rho_0}=-\frac{T_1}{T_0}\,,
\eeq
i.e. in the axially symmetric waves considered here the density 
variations are in counter-phase with temperature variations.

The viscous dissipative function $Q_{visc}$ can be written as $Q_{visc}=\rho\nu\Phi$, where the function $\Phi$ in polar coordinates is
\begin{eqnarray}
\Phi=&2\left[
\myfrac{\partial u_r}{\partial r}^2+
\left(
\frac{1}{r}\myfrac{\partial u_\phi}{\partial \phi} +\frac{u_r}{r}
\right)^2+ 
\myfrac{\partial u_z}{\partial z}^2
 \right] \nonumber\\
&+\left[
r\frac{\partial}{\partial r}\myfrac{u_\phi}{r}+
\frac{1}{r}\frac{\partial u_r}{\partial \phi}\right]^2 +
 \left[\frac{1}{r}\frac{\partial u_z}{\partial \phi}\right]^2
 \nonumber\\
&+\left[\frac{\partial u_r}{\partial z} +
\frac{\partial u_z}{\partial r}\right]^2 -
\frac{2}{3}(\nabla\cdot\bm{u})^2\,.
 \nonumber\\
\end{eqnarray} 
All terms but one in this function are quadratic
in small velocity perturbations; this term has the form:
\beq{qterm}
\nu\rho \left( \frac{\partial u_\phi}{\partial r}-\frac{u_\phi}{r}\right)^2\,.
\eeq 
Writing for the azimuthal velocity $u_\phi=u_{\phi,0}+u_{\phi,1}$ (here for the purposes of this paragraph 
and only here we specially mark the unperturbed velocity with index 0, 
not to be confused with our notations $u_\phi$ for perturbed velocity in \Eq{iur}-\Eq{iuphi} above and below). 
Thus we obtain for the viscous dissipation
\beq{Qv}  
Q_{visc}=\nu\rho r \frac{d \Omega}{d r}\left[r\frac{d\Omega}{d
r}-2ik_r u_{\phi,1}-2\frac{u_{\phi,1}}{r}\right] + \hbox{quadratic\, terms}\,.
\eeq 
Here $\Omega=u_{\phi,0}/r$ is the angular (Keplerian) velocity of the unperturbed flow. 
The first term in parentheses describes the viscous energy release in the unperturbed Keplerian flow. For this unperturbed flow we have
\beq{}
\frac{\partial s_0}{\partial t}=\nu\mu \frac{[r(d \Omega/d r)]^2}{{\cal R}T}=\frac{9}{4}\nu\mu\frac{\Omega^2}{{\cal R}T}\,.
\eeq
Thus, the entropy of the unperturbed flow changes along the radius. However, on the scale of the order of or smaller than the disc thickness, the entropy gradient can be neglected.  
The second term in \Eq{Qv} vanishes if $k_r=0$, i.e.  
we consider two-dimensional perturbations with only $k_z\ne 0$. 
As a result, the energy equation with zero entropy gradients in the
Boussinesq limit becomes
\beq{e:en1}
\frac{\rho_0{\cal R} T_0}{\mu}s_1=-2ik_r \nu\rho_0 r \frac{d \Omega}{d r}u_{\phi,1}
-\kappa_T k^2 T_0\frac{T_1}{T_0}\,.
\eeq
Like in the linearized equation $\nabla\cdot \bm{u}=0$, here we have neglected the term $u_{\phi,1}/r$.
By substituting \Eq{e:s1} 
and \Eq{e:T1} into \Eq{e:en1}, we find the relation between the density variations and
$u_\phi$ in the in the Boussinesq limit with zero entropy gradients: 
\beq{e:rho1uphi}
\frac{\rho_1}{\rho_0}\left(i\omega c_p + \frac{\kappa_T k^2}{\rho_0{\cal R}/\mu}\right)=
\frac{2ik_r \nu r (d \Omega/d r)}{{\cal R}T_0/\mu}u_{\phi}
\eeq 
Here $c_p=\gamma c_V=\gamma/(\gamma-1)$ is the specific heat capacity (per particle) at constant pressure. 

To describe the effects of thermal conductivity, 
it is convenient to introduce the usual dimensionless Prandtl number:
\beq{e:Pre}
\mathrm{Pr}\equiv \frac{\nu\rho_0 C_p}{\kappa_T}\,.
\eeq
(Here $C_p=c_p{\cal R}/\mu$).
Substituting \Eq{e:Pre} into \Eq{e:rho1uphi} yields:
\beq{e:r1ufi}
\frac{\rho_1}{\rho_0}=\frac{\gamma/(\gamma-1)}{(i\omega+\nu k^2/\mathrm{Pr})}
\frac{2ik_r \nu r (d \Omega/d r)}{{\cal R}T_0/\mu}u_{\phi}
\eeq

It is straightforward to include the density perturbations 
in the non-adiabatic case \eqn{e:rho1uphi} 
in the analysis. This significantly complicates the final dispersion equation 
(see \Eq{e:eq*gen} below). We stress again that the two-dimensional case with $k_r=0$ 
produces the dispersion relation for small local perturbations which is \textit{exact} 
even in the case of non-adiabatic perturbations.

c) The three components of the induction equation with account for 
$\eta\Delta \bm{B}\to -\eta k^2\{ b_r,b_\phi,b_z\}$ read:
\beq{ibr}
i\omega b_r=-iB_0k_zu_r-\eta k^2 b_r\,,
\eeq
\beq{ibphi} 
i\omega b_\phi=-iB_0k_zu_\phi+r\frac{d\Omega}{dr}b_r-\eta k^2b_\phi\,,
\eeq
\beq{ibz}
i\omega b_z=iB_0k_ru_r-\eta k^2b_z\,.
\eeq

Following \cite{1998bhad.conf.....K}, we express all perturbed quantities
through $u_z$:
\beq{ur}
u_r=-\frac{k_z}{k_r}u_z\,,
\eeq
\beq{uphi}
u_\phi=\frac{k_z}{k_r}
\frac{\frac{\kappa^2}{2\Omega}(i\omega+\eta k^2)^2+c_A^2k_z^2r\frac{d\Omega}{dr}}
{\left[(i\omega+\nu k^2)(i\omega+\eta k^2)+c_A^2k_z^2\right](i\omega+\eta k^2)}u_z\,,
\eeq
\beq{br}
\frac{b_r}{B_0}(i\omega+\eta k^2)=i\frac{k_z^2}{k_r}u_z\,,
\eeq
\beq{bphi}
\frac{b_\phi}{B_0}(i\omega+\eta k^2)=-ik_zu_\phi+\frac{ir\frac{d\Omega}{dr}}{(i\omega+\eta k^2)}\frac{k_z^2}{k_r}u_z\,,
\eeq
\beq{bz}
\frac{b_z}{B_0}(i\omega+\eta k^2)=-i k_z u_z\,,
\eeq

The system of linear equations \eqn{Bq} and \eqn{ur}-\eqn{bz} contains the 
equation $\nabla\cdot\bm{b}=0$. Indeed, by multiplying \Eq{br} and \Eq{bz} by $k_r$ and  $k_z$,
respectively, and summing up the obtained equations, we get $k_rb_r+k_zb_z=0$.
Substituting \Eq{ur}-\Eq{bz} into \Eq{iur} and rearranging the terms, we arrive at the
dispersion relation \eqn{e:eq*}.

The dispersion relation 
in the general case of non-adiabatic perturbations with $k_r\ne 0$, i.e. 
with non-vanishing density perturbations $\rho_1$ (see \Eq{e:rho1uphi}) is:
\begin{eqnarray}
\label{e:eq*gen}
\omega_{**}^4+\myfrac{k_z}{k}^2 & \left[\left(i\omega+\eta k^2\right)^2\kappa^2+c_A^2k_z^2
(\kappa^2-4\Omega^2)\right] \nonumber \\
&\left[1-\frac{\gamma-1}{\gamma}
\frac{ik_r}{(i\omega+\nu k^2/\mathrm{Pr})}\left(A-\frac{k_r}{k_z}B\right)\right]=0\,,
\end{eqnarray}
where $\omega_{**}$ is determined by \Eq{e:omega**} in the main text and  
\beq{}
A=\nu
\left(\frac{d\ln\Omega}{d\ln r}\right)\left(\frac{1}{p_0}\frac{dp_0}{dr}\right)\,;
\quad B=\nu
\left(\frac{d\ln\Omega}{d\ln r}\right)\left(\frac{1}{p_0}\frac{dp_0}{dz}\right)\,.
\eeq
Although the terms with $A$ and $B$ arising from the viscous dissipation function are proportional to $(k_r/r)(\nu/\omega)$ and $(k_r^2/k_z r)(\nu/\omega)$, they are retained in our analysis because at large viscosity they can be comparable to or even higher than one.  
The expression in the square brackets in \Eq{e:eq*gen} above can be 
rewritten in the equivalent form:
\beq{}
\left[1+\frac{\gamma-1}{\gamma}
\frac{i\nu}{(i\omega+\nu k^2/\mathrm{Pr})}\myfrac{k_r}{k_z}\frac{d\ln\Omega/d\ln r}{{\cal R}T_0/\mu}
(k_zg_{r,eff}-k_rg_z)\right]\,,
\eeq
where $g_{r,eff}=-1/\rho_0 (dp_0/dr)$ and $g_z=-1/\rho_0 (dp_0/dz)$ are the effective radial and 
vertical gravity accelerations in the unperturbed flow, respectively.
Clearly, for $k_r=0$ we return to \Eq{e:eq*} with $k=k_z$. 
Note that for $k_r\ne 0$
\Eq{e:eq*gen} is a fifth-order algebraic equation. For perturbations with $k_r=0$ 
this equation becomes a fourth-order algebraic equation, which already has exponentially 
growing MRI modes. For completeness, it would be desirable to investigate
this five-order equation. However, in the absence of the magnetic field \Eq{e:eq*gen}
turns into a third-order algebraic equation. As we show in the subsequent paper
(Shakura \& Postnov 2014, submitted), one of the Rayleigh modes in this 
case becomes exponentially unstable at long wavelengths even in the Rayleigh-stable 
case of Keplerian rotation.

\section{Analytical solution for the critical wave number $\tilde k_{cr}$ in the general case of 
non-ideal plasma}
\label{A:analyt}

Here we provide the analytical solution of \Eq{e:eq*} for arbitrary magnetic Prandtl number 
$\mathrm{P_m}$ and dimensionless viscosity coefficient $\tilde \nu$ at the neutral point where $\tilde \omega(\tilde k_{cr})=0$. To do this, it is convenient, for the sake of brevity, 
to introduce new dimensionless variables
\beq{}
y\equiv \tilde k^2, \quad X=i\tilde\omega +\tilde\nu y
\eeq
and rewrite dimensionless dispersion relation \eqn{e:eq*} in the equivalent form:
\begin{eqnarray}
& X^4+2\frac{1-\mathrm{P_m}}{\mathrm{P_m}}\tilde\nu y X^3+
\left[\myfrac{1-\mathrm{P_m}}{\mathrm{P_m}}^2
\tilde\nu^2 y^2+2y+1\right]X^2+ \nonumber\\
&\left[\frac{1-\mathrm{P_m}}{\mathrm{P_m}}\tilde\nu y(y+1)\right]X+
\myfrac{1-\mathrm{P_m}}{\mathrm{P_m}}^2
\tilde\nu^2 y^2 + y^2-3y=0\,.
\end{eqnarray}
(Here we assumed Keplerian discs with $\tilde \kappa=1$ and used $k_z/k=1$).
Noticing that at the neutral point determined by the condition $\tilde\omega(y_{cr})=0$ 
we have $X=\tilde \nu y_{cr}$, we arrive at the equation for $y_{cr}$:
\beq{e:cubic}
y_{cr}\left[\tilde \nu^4 y_{cr}^3+\tilde\nu^2 y_{cr}(2y_{cr}\mathrm{P_m}+1)+\mathrm{P_m}^2(y_{cr}-3)\right]=0\,.
\eeq
At $\mathrm{P_m}=1$ this equation, of course, coincides with \Eq{e:kcrPm1}.
The non-trivial real solution to the cubic equation in the square brackets of 
\Eq{e:cubic} reads: 
\beq{e:ycr}
y_{cr}\equiv \tilde k_{cr}^2={\cal A}-\frac{2\mathrm{P_m}}{3\tilde\nu^2}-
\frac{1}{\cal A}\left(\frac{1}{3\tilde\nu^2}-\frac{\mathrm{P_m}^2}{9\tilde\nu^4}
\right)\,,
\eeq
where
\begin{eqnarray}
{\cal A}=&\left[
\left(
\frac{1}{27\tilde\nu^6}+
\frac{2\mathrm{P_m}^2}{27\tilde\nu^8}+
\frac{\mathrm{P_m}^3}{\tilde\nu^8}+ 
\frac{9\mathrm{P_m}^4}{4\tilde\nu^8}+
\frac{\mathrm{P_m}^4}{27\tilde\nu^{10}}+
\frac{\mathrm{P_m}^5}{9\tilde\nu^{10}}
\right)^{1/2}+\right. \nonumber \\
& \left.\frac{\mathrm{P_m}}{3\tilde\nu^4}+
\frac{3\mathrm{P_m}^2}{2\tilde\nu^4}+
\frac{\mathrm{P_m}^3}{27\tilde\nu^6}
\right]^{1/3}\,.
\end{eqnarray} 
At high dimensionless viscosities there is an asymptotic to the solution \eqn{e:ycr} 
for $ \mathrm{P_m}/\tilde\nu^2\ll 1$:
\beq{e:kcrasa}
y_{cr}=\tilde k_{cr}^2\approx \frac{3\mathrm{P_m}^2/\tilde\nu^2}{1+\mathrm{P_m}^2/\tilde\nu^2}=
\frac{3\mathrm{P_m}^2}{\tilde\nu^2}+{\cal O}\myfrac{\mathrm{P_m}^2}{\tilde\nu^2}^2\,.
\eeq
Note that this asymptotic can be also 
found in \cite{2008ApJ...684..498P} (their Eq. (97))
and for small $\mathrm{P_m}$ can be derived for Keplerian rotation and $k=k_z$ 
from Eq. (3) in \cite{2001MNRAS.325L...1J}.

\bibliographystyle{mn2e}
\expandafter\ifx\csname natexlab\endcsname\relax\def\natexlab#1{#1}\fi
\bibliography{mri1}

\begin{thebibliography}{31}
\expandafter\ifx\csname natexlab\endcsname\relax\def\natexlab#1{#1}\fi

\bibitem[{{Acheson}(1978)}]{1978RSPTA.289..459A}
{Acheson} D.~J., 1978, Royal Society of London Philosophical Transactions
  Series A, 289, 459

\bibitem[{{Balbus}(2004)}]{2004ApJ...616..857B}
{Balbus} S.~A., 2004, \apj, 616, 857

\bibitem[{{Balbus}(2012)}]{2012MNRAS.423L..50B}
{Balbus} S.~A., 2012, \mnras, 423, L50

\bibitem[{{Balbus} \& {Hawley}(1991)}]{1991ApJ...376..214B}
{Balbus} S.~A., {Hawley} J.~F., 1991, \apj, 376, 214

\bibitem[{{Balbus} \& {Hawley}(1998)}]{1998RvMP...70....1B}
{Balbus} S.~A., {Hawley} J.~F., 1998, Reviews of Modern Physics, 70, 1

\bibitem[{{Balbus} \& {Henri}(2008)}]{2008ApJ...674..408B}
{Balbus} S.~A., {Henri} P., 2008, \apj, 674, 408

\bibitem[{{Chandrasekhar}(1960)}]{1960PNAS...46..253C}
{Chandrasekhar} S., 1960, Proceedings of the National Academy of Science, 46,
  253

\bibitem[{{Hawley}, {Gammie} \& {Balbus}(1995){Hawley}, {Gammie}, \&
  {Balbus}}]{1995ApJ...440..742H}
{Hawley} J.~F., {Gammie} C.~F., {Balbus} S.~A., 1995, \apj, 440, 742

\bibitem[{{Hawley} {et~al}\mbox{.}(2013){Hawley}, {Richers}, {Guan}, \&
  {Krolik}}]{2013ApJ...772..102H}
{Hawley} J.~F., {Richers} S.~A., {Guan} X., {Krolik} J.~H., 2013, \apj, 772,
  102

\bibitem[{{Islam} \& {Balbus}(2005)}]{2005ApJ...633..328I}
{Islam} T., {Balbus} S., 2005, \apj, 633, 328

\bibitem[{{Ji}, {Goodman} \& {Kageyama}(2001){Ji}, {Goodman}, \&
  {Kageyama}}]{2001MNRAS.325L...1J}
{Ji} H., {Goodman} J., {Kageyama} A., 2001, \mnras, 325, L1

\bibitem[{{Kato}, {Fukue} \& {Mineshige}(1998){Kato}, {Fukue}, \&
  {Mineshige}}]{1998bhad.conf.....K}
{Kato} S., {Fukue} J., {Mineshige} S., eds., 1998, {Black-hole accretion disks}

\bibitem[{{Ketsaris} \& {Shakura}(1998)}]{1998A&AT...15..193K}
{Ketsaris} N.~A., {Shakura} N.~I., 1998, Astronomical and Astrophysical
  Transactions, 15, 193

\bibitem[{{Kirillov}, {Stefani} \& {Fukumoto}(2014){Kirillov}, {Stefani}, \&
  {Fukumoto}}]{2014JFM...760..591K}
{Kirillov} O.~N., {Stefani} F., {Fukumoto} Y., 2014, Journal of Fluid
  Mechanics, 760, 591

\bibitem[{{Kotko} \& {Lasota}(2012)}]{2012A&A...545A.115K}
{Kotko} I., {Lasota} J.-P., 2012, \aap, 545, A115

\bibitem[{Kundu, Cohen \& Dowling(2012)Kundu, Cohen, \& Dowling}]{Kundu5ed}
Kundu P.~K., Cohen I.~M., Dowling D.~R., 2012, {Fluid Mechanics}, 5th edn.
  Academic Press, Boston

\bibitem[{{Lord Rayleigh}(1916)}]{LordRayleigh16}
{Lord Rayleigh}, 1916, Proc. R. Soc. A, 93, 143

\bibitem[{{Masada} \& {Sano}(2008)}]{2008ApJ...689.1234M}
{Masada} Y., {Sano} T., 2008, \apj, 689, 1234

\bibitem[{{Nauman} \& {Blackman}(2015)}]{2014arXiv1409.2442N}
{Nauman} F., {Blackman} E.~G., 2015, \mnras, 446, 2102

\bibitem[{{Pessah} \& {Chan}(2008)}]{2008ApJ...684..498P}
{Pessah} M.~E., {Chan} C.-k., 2008, \apj, 684, 498

\bibitem[{{Ruediger} {et~al}\mbox{.}(2014){Ruediger}, {Schultz}, {Stefani}, \&
  {Mond}}]{2014arXiv1407.0240R}
{Ruediger} G., {Schultz} M., {Stefani} F., {Mond} M., 2014, ArXiv e-prints
  1407.0240

\bibitem[{{Sano} \& {Miyama}(1999)}]{1999ApJ...515..776S}
{Sano} T., {Miyama} S.~M., 1999, \apj, 515, 776

\bibitem[{{Shakura} \& {Sunyaev}(1973)}]{1973A&A....24..337S}
{Shakura} N.~I., {Sunyaev} R.~A., 1973, \aap, 24, 337

\bibitem[{{Shirakawa} \& {Hoshino}(2014)}]{2014PhPl...21e2903S}
{Shirakawa} K., {Hoshino} M., 2014, Physics of Plasmas, 21, 052903

\bibitem[{{Sorathia} {et~al}\mbox{.}(2012){Sorathia}, {Reynolds}, {Stone}, \&
  {Beckwith}}]{2012ApJ...749..189S}
{Sorathia} K.~A., {Reynolds} C.~S., {Stone} J.~M., {Beckwith} K., 2012, \apj,
  749, 189

\bibitem[{{Spiegel} \& {Veronis}(1960)}]{1960ApJ...131..442S}
{Spiegel} E.~A., {Veronis} G., 1960, \apj, 131, 442

\bibitem[{{Spitzer}(1962)}]{1962pfig.book.....S}
{Spitzer} L., 1962, {Physics of Fully Ionized Gases}

\bibitem[{{Stone}(2011)}]{2011IAUS..274..422S}
{Stone} J.~M., 2011, in IAU Symposium, Vol. 274, IAU Symposium, {Bonanno} A.,
  {de Gouveia Dal Pino} E., {Kosovichev} A.~G., eds., pp. 422--428

\bibitem[{{Suleimanov}, {Lipunova} \& {Shakura}(2008){Suleimanov}, {Lipunova},
  \& {Shakura}}]{2008A&A...491..267S}
{Suleimanov} V.~F., {Lipunova} G.~V., {Shakura} N.~I., 2008, \aap, 491, 267

\bibitem[{{Suzuki} \& {Inutsuka}(2014)}]{2014ApJ...784..121S}
{Suzuki} T.~K., {Inutsuka} S.-i., 2014, \apj, 784, 121

\bibitem[{{Velikhov}(1959)}]{Velikhov59}
{Velikhov} E.~P., 1959, Sov. Phys. JETP, 36, 1398

\end{thebibliography}

\end{document}